\newcommand\etal{et~al.}
\newcommand\kms{\ifmmode {\rm\,km\,s^{-1}}\else${\rm\,km\,s^{-1}}$\fi}
\newcommand\simlt{\mathrel{\spose{\lower 3pt\hbox{$\mathchar"218$}} \raise 2.0pt\hbox{$\mathchar"13C$}}}
\newcommand\simgt{\mathrel{\spose{\lower 3pt\hbox{$\mathchar"218$}} \raise 2.0pt\hbox{$\mathchar"13E$}}}
\newcommand\aap{{\em A\&A}}

\newcommand\aaps{{\em A\&AS}}
\newcommand\aj{{\em AJ}}
\newcommand\apj{{\em ApJ}}
\newcommand\apjl{{\em ApJ}}

\newcommand\mnras{{\em MNRAS}}

\documentclass[usenatbib]{mn2e}

\usepackage[dvips]{graphicx}

\begin{document}

\title[The DRGs population of QSOs and obscured sources]{The distant red galaxy neighbour population of $1\leq z \leq 2$ QSOs and optically obscured sources\thanks{Based on
observations and/or data products by the Flamingos Extragalactic Survey.
 FLAMINGOS was designed and constructed by the IR instrumentation group 
(PI: R. Elston) at the University of Florida, Department of Astronomy, 
with support from NSF grant AST97-31180 and Kitt Peak National Observatory 
}}

\author[C. Bornancini \etal]{
\parbox[t]{\textwidth}{
Carlos~G.~Bornancini$^{1,2}$, Diego~G.~Lambas$^{1,3}$
}
\vspace*{6pt} \\ 
$^1$Grupo de Investigaciones en Astronom\'\i a Te\'orica y Experimental, IATE\\
Observatorio Astron\'omico, Universidad Nacional de C\'ordoba\\
Laprida 854, X5000BGR, C\'ordoba, Argentina.\\
$^2$ Secretar\'ia de Ciencia y T\'ecnica de la Universidad Nacional de C\'ordoba.\\
$^3$Consejo Nacional de Investigaciones Cient\'\i ficas y T\'ecnicas (CONICET),
Avenida Rivadavia 1917, C1033AAJ, Buenos Aires, Argentina.\\
}
\pubyear{2007}

\maketitle


\begin{abstract}
We study the Distant Red Galaxy (DRG, $J-K_s>2.3$) neighbour population of Quasi Stellar Objects (QSOs) selected from the Sloan Digital Sky Survey (SDSS) in the redshift range  $1\leq z\leq 2$. We perform a similar analysis for optically obscured AGNs (i.e. with a limiting magnitude $I > 24$) detected in the mid-infrared (24 $\mu$m) with the Spitzer Space Telescope and a mean redshift $z\sim 2.2$ in the Flamingos Extragalactic Survey (FLAMEX). Both QSOs and obscured AGN target samples cover 4.7 deg$^2$ in the same region of the sky. 
We find a significant difference in the environment of these two target samples.
Neighbouring galaxies close to QSOs tend to be bluer than galaxies in optically obscured source environments. 
We also present results on the cross--correlation function of DRGs around QSOs and optically faint mid-infrared sources.
The corresponding correlation length obtained for the QSO sample targets is $r_0$=$5.4\pm1.6$
 Mpc h$^{-1}$ and a slope of $\gamma$=$1.94\pm0.10$ . For the optically obscured galaxy sample we find  $r_0$=$8.9\pm1.4$ Mpc h$^{-1}$ and a slope of $\gamma$=$2.27\pm0.20$.
These results indicate that optically faint obscured sources are located in denser environment of evolved red galaxies compare to QSOs. 
 
\end{abstract}


\section{Introduction}

Studies of large samples of distant galaxies are fundamental to provide a deeper insight in the formation and evolution of galaxies and systems of galaxies, such as clusters and groups. 
Using measurements of galaxy clustering at $z\sim 1-2$ one can test the predictions of cosmological models of structure formation and evolution \citep{kauff}. 
In the last years diverse photometric selection techniques have been developed in order to select high redshift galaxies.
In particular the Lyman break technique is an ideal method for selecting a large number of distant galaxies and for studying the large-scale distribution and properties of star-forming systems at high redshifts from multicolour optical data \citep{steidel96, madau} . This technique requires a high rest-frame far ultraviolet luminosity, and so will preferentially select galaxies with recent or unobscured active star-formation \citep{steidel99}. A new near-infrared selection technique has been developed in recent years to select samples of galaxies at high redshifts.
The Distant Red Galaxies (DRGs, hereafter) colour--cut criterion ($J-K_s>2.3$, Vega system) \citep{saracco,franx,van} is expected to select galaxies with prominent rest frame optical breaks, caused by the 3625 \AA~ Balmer-break or the  4000 \AA ~Ca II H+K break.
The Balmer discontinuity at 3625 \AA~ is strongest in A-type stars and the 4000 \AA~ break is characteristic of coolers stars with types later than G0 and strong in giant and supergiant stars \citep{forster}.
van Dokkum et al. (2003) found that the DRG criterion selects galaxies with rest-frame optical colours similar to those of normal nearby galaxies. Using deep mid-infrared observations with IRAC on the Spitzer Space Telescope, \citet{labbe} found that 70\% of the DRGs are best described by dust-reddened star forming models and 30\% are very well fit with old stellar population, passive star evolution galaxy models. By comparison of the stellar populations of DGRs to those obtain in Lyman-break galaxies (LBGs), \citet{labbe} found that the average mass-to light ratios ($M/L_K$) of the DRGs are about three times higher than the LBGs sample, indicating that DGRs may represent massive and old galaxies, similar to those found in the local Universe. A similar study by \citet{forster} comparing LBGs and DRGs at similar redshifts and rest-frame $V$-band luminosities, shows that DRGs are older, more massive and more obscured for any given star formation history than LBGs. 
Recent works by \citet{conse} and \citet{graziana} show that this single near IR colour--cut selects a rather heterogeneous sample of galaxies, from distant luminous massive systems, to a significant fraction of less luminous dusty star-burst and galaxies with mixed morphology at redshifts $z\sim2$, with extended tails at $z=1$ and $z=4$.

At low redshifts, QSO environments show similar characteristics of those found in normal galaxies \citep{smith,coldwell}. Moreover, \citet{smith} found that the cross-correlation function between low ($z < 0.3$) QSOs and galaxies, is consistent with the auto-correlation function of galaxies selected from the APM Galaxy Survey. Unlike radio-loud QSOs, \citet{elli} found that radio-quiet QSOs at $0.3<z<0.6$ are rarely found in high density environments as rich as Abell class 1. 
Recently, \citet{serber} studying the environment of $z<0.4$ luminous ($M_{i}<-$22) quasars in the SDSS area, found that they are located in higher local overdensity regions than are typical $L^*$ galaxies.
However, the results for QSO environment at higher redshifts ($1\leq z\leq 2$) are contradictory. \citet{yee} found that some of the $z\sim0.6$ QSOs are located in environments as rich as those of Abell class 1 clusters. \citet{hall} reported a significant excess of faint galaxies in the fields of  $z=1-2$ quasars. No excess galaxy population associated with radio-quiet QSOs  are reported by \citet{boyle} at $z\sim1$. \citet{croom} found an anti-correlation between radio-quiet QSOs and galaxies at redshifts $z\sim 1-1.5$.  
\citet{coil} studied the clustering of galaxies around a sample of $0.7< z<1.4$ QSOs selected from the SDSS and DEEP2 surveys. They found, from a two-point cross-correlation analysis, that the local environment of QSOs is consistent with the mean environment of the full DEEP2 galaxy population and that they cluster similar to the blue, star-forming galaxies rather than the red galaxies. Their results imply that high redshift QSOs do not reside in particularly massive dark matter halos.

Using the DRGs criterion, \citet{kaji} reported a discovery of proto-clusters candidates around 6 high redshifts radio galaxies at $z\sim2.5$ on the basis of excess of $J-K_s$ colour--selected galaxies.
Recently, a similar colour criterion was adopted to identify and to study the environment of very distant objects. This method uses a colour cut $i_{775}-z_{850}>1.3$ which select objects at $z\sim6$. An over-density of galaxies around the most distant radio--loud QSO (SDSS J0836+0054 at $z=5.8$) was reported by \citet{wei}, who found a surface density six times higher than the number expected from deep fields at similar redshifts. 
 \citet{stiavelli} found an excess of $i_{775}-z_{850}>1.3$ galaxies around one of the most distant QSOs at $z=6.28$.

In this paper we investigate the population and colour distribution of DRGs, selected at bright near--IR magnitudes ($K_s<19.5$, Vega system) around high redshift QSOs ($1\leq z\leq 2$ ) and optical obscured sources ($z\sim2.2$) in the FLAMINGOS Extragalactic Survey (FLAMEX, \citet{elston}).  

This paper is organised as follows: Section 2 describes the sample analysed, 
The DRG number counts are analysed in Section 3. 
In Section 4 we study the colour distribution of galaxies in different environments.
We investigate the QSO--DRGs and Obscured--DRGs cross-correlation 
analysis in Section 5.
Finally we discuss our results in Section 6.

In this work we assume a standard $\Lambda$CDM model Universe with cosmological parameters, $\Omega_{M}$=0.3, $\Omega_{\Lambda}$=0.7 and a Hubble constant of $H_0=$100 Km~s$^{-1}$Mpc$^{-1}$. All magnitudes are expressed in the Vega system.

\section{Observational Sample}
The FLAMINGOS Extragalactic Survey (FLAMEX, \citet{elston}) is a wide-area, 
near-infrared imaging survey in $J$ and $K_s$ bands within the NOAO Deep 
Wide-Field Survey (NDWFS) regions \citep{jannu}.
This paper uses catalogues of the first Data Release DR1 of the northern part of the Survey (Bo{\"o}tes field) that covers 4.7 deg$^2$ in both $J$ and $K_s$ bands \footnote{http://flamingos.astro.ufl.edu/extragalactic/}. 
Using Monte Carlo simulations \citet{elston} found that more than 90 \% of the survey region is complete to $K_s=$ 19.2, with 50 \% being complete to $K_s=19.5$. A detailed description of the observing strategy and data reduction can be found in \citet{elston}.   
The catalogues were made using $K_s$--selected objects for each survey subset with the SExtractor package \citep{bertin}, using dual image mode to measure the 
$J$--band magnitudes within the same regions. Detections in the different bands were matched if the centroids were within 1 $\arcsec$ of each other. 
$J-K_s$ colours were calculated using 4$\arcsec$ diameter aperture photometry, 
sufficiently large to avoid intrinsic aberrations that cause the PSF to vary 
significantly across the field and to produce robust 
photometry across entire subfields. 

In order to reject spurious objects lying in the edges of the images,
we selected objects that not overlap with another object 
using the SExtractor parameters ${\tt {FLAGS=0}}$ and ${\tt {WEIGHT\_MAP}}$ $<$ 0.7, in both $J$ and $K_s$ bands.

For two different reasons, no star-galaxy separation was performed. First, the variable PSF in the FLAMEX area precluded use of structural information to separate stars from extended sources and second, DRGs are of such small apparent size that considerable fraction of them would possibly be misclassified as stars and rejected from the final catalogues.
As noted by \citet{elston}, and as can be seen in the colour-magnitude diagram (Figure 1), the $J-K_s$ colours alone provides a simple means of performing star-galaxy separation.
The two horizontal sequences at $J-K_s$=0.4 and $J-K_s$=0.8 correspond to galactic stars with types later than G5 and earlier than K5, respectively \citep{finlator}.
Red galaxies with $J-K_s>$2.3 are well separated from stars.
We also show in Figure 1 the DRGs criterion adopted (Dashed lines).
Our final catalogue consist of 7131 DRGs in a contiguous area of 4.7 degree$^2$, representing the largest sample of DRGs selected at bright magnitudes $K_s<19.5$. 

\begin{figure}
\includegraphics[width=80mm]{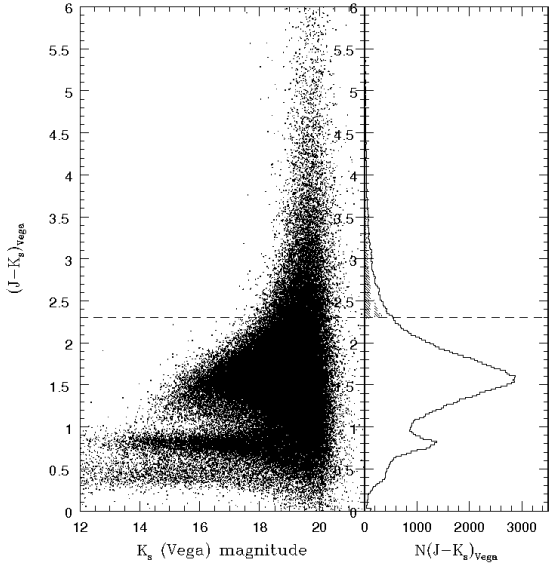}
\caption{Left panel: $J-K_s>$ vs $K_s$ (Vega) colour-magnitude diagram for the 
FLAMEX field (Bo{\"o}tes area). 
Right panel: Colour distribution. Dashed lines show the DRGs criterion adopted.}
\label{ic}
\end{figure}

The QSO sample analysed in this work was obtained using the NED Database. \footnote{http://nedwww.ipac.caltech.edu/ - the NASA-IPAC Extragalactic Database}
We selected QSOs in the redshift range  $1\leq z\leq 2$ in the FLAMEX area. The sample consists of twelve QSOs selected from the Sloan Digital Sky Survey (SDSS, Data Release 1 to 4) \citep{york,richards} and one X--ray selected AGN \citep{puch} (See Table 2).

We also choose a sample of optically obscured galaxies discovered with the Spitzer Space Telescope, which are either optically very faint ($R>$24.5) or invisible ($R>$26) with no counterparts in the NOAO Deep Wide-Field Survey regions. The main goal of this survey is to select optically very faint galaxies but bright enough in the mid-infrared for redshift determination with the Infrared Spectrograph on Spitzer (IRS), which would not have been identified in previous optical studies. Sources with $I > 24$ mag were identified with the Multi-band Imaging Photometer (MIPS) on the Spitzer Space Telescope to a limiting 24 $\mu$m flux density above 0.75 mJy. 114 sources of the total sample met this optical criterion. Out of these, 17 sources have no optical or have very faint optical counterparts. Rejecting sources lying in the edges of the images, where noise level are high and signal to noise values are low, we finally selected 6 sources located within the FLAMEX area, with  
a mean redshift distribution $z\sim2.2$. The redshift range of the selected sources are similar to that of the redshift distribution of DRGs with $K_s<20.15$ (Vega magnitudes) found by \citet{graziana}. Redshifts from this sample were derived primarily from strong silicate absorption features \citep{houck}. We also selected another two sources with feasible redshift determinations derived by a weak silicate absorption feature, based on fits of redshifted spectral templates of local Ultra-luminous Infrared Galaxies (ULIRGS) from \citet{weedman}.
Full details of the sample selection and redshift determination is given in \citet{houck, weedman}.

As noted by \citet{houck} template fits suggest that most of the optically obscured sources are dominated by an AGN component, i.e. a steeply rising spectra and varying levels of silicate absorption, similar to the infrared spectra of known local AGNs. 
By comparing bolometric luminosities of submillimeter selected galaxies, such as those detected in SCUBA or MAMBO suryeys, \citet{houck} found that the obscured source sample seems to represent sources that have a hotter dust component than typical submillimeter-selected galaxies, assuming that mid-infrared luminosities scale to bolometric luminosities similar than in template sources which fit the mid-infrared spectra. Using this scaling, the obscured sources have implied infrared luminosities between 6$\times10^{12}$ L$_{\sun}$ and 6$\times10^{13}$ L$_{\sun}$, similar to bright QSOs at $z\sim2$ \citep{hop}. 

\begin{figure}
\includegraphics[width=80mm]{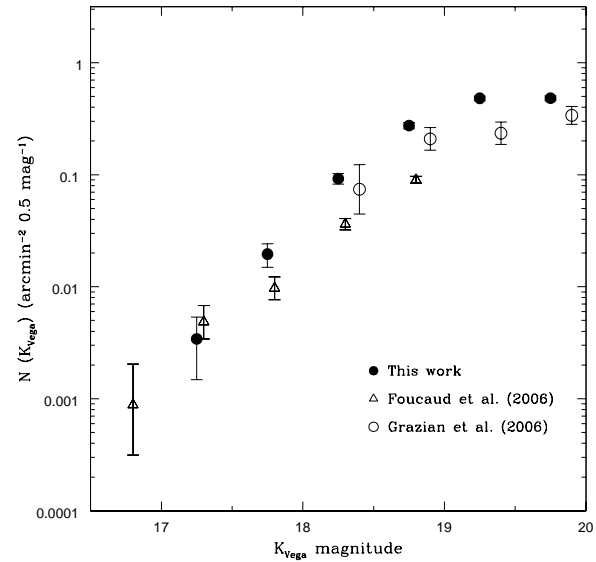}
\caption{Differential number counts as a function of $K_s$ magnitude for the DRG sample (Filled circles). The error bars plotted represent Poissonian errors (3 $\sigma$). We include the DRG number counts obtained in the literature. Open triangles represent data from the UKIDSS Ultra Deep Survey Early Data Release (Foucaud et al. 2006, private communication) and open circles are data obtained in the GOODS-MUSIC Sample, \citet{graziana}.}
\label{ic}
\end{figure}

\section{Galaxy number counts}

We have computed the differential number counts of DRGs per unit area as a function of $K_s$ magnitude which are shown in Figure 2. We use SExtractor ${\tt {MAG\_AUTO}}$ magnitudes for this calculation and estimate error bars using 3 $\sigma$ uncertainties 
estimated using Poissonian errors of the raw galaxy counts (See Table 1).
We compare our determinations with $K-$band number counts for DRGs obtained by 
\citet{graziana} in the GOODS--MUSIC Sample and with estimates obtained in the UKIDSS Ultra Deep Survey Early Data Release (Foucaud et al. 2006, private communication).
We have done this comparison using Vega magnitudes, assuming a transformation $K_{Vega}=K_{AB}-1.85$ \citep{cool}.
For $K_s>$17.5 our counts are systematically higher than those obtain by Foucaud et al. This discrepancy can be explained by cosmic variance because of the small area (0.62 deg$^2$) of the Early Data Release of the UKIDSS Ultra Deep Survey. \\

\bigskip
\setcounter{table}{0}
\label{t0}
Table 1. DRG number counts as a function of $K_s$ band.
The columns give the central bin magnitude, the raw counts, counts in units of arcmin$^{-2}$mag$^{-1}$ and 3 $\sigma$ uncertainties 
estimated using Poissonian errors.
\begin{tabular}{lrcl}
\hline
(1)&(2)&(3)&(4)\\
$bin$&$N$ &log($\Sigma$)& Error (3$\sigma$)\\
& &\small{arcmin$^{-2}$mag$^{-1}$}&  \\
\hline
17.25&   28  &$-$2.466 &   0.0019 \\
17.75&  160  &$-$1.709 &   0.0046 \\
18.25&  756  &$-$1.035 &   0.0100 \\
18.75& 2248  &$-$0.561 &   0.0173 \\
19.25& 3939  &$-$0.318 &   0.0229 \\
19.75& 3944  &$-$0.317 &   0.0230 \\
\hline
\end{tabular}
 
\begin{figure}
\includegraphics[width=80mm]{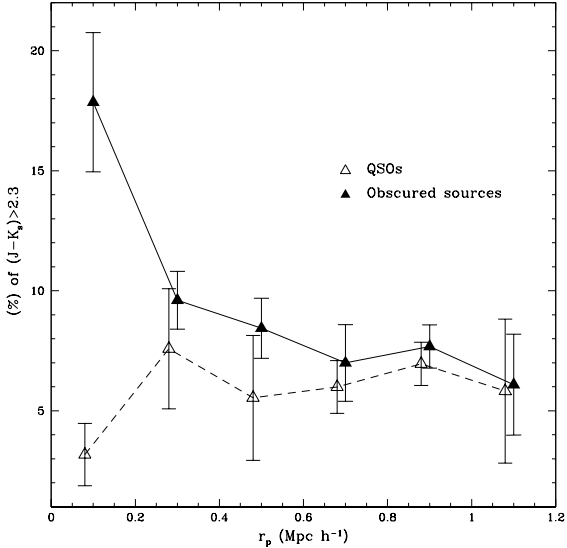}
\caption{Percentage of galaxies with $J-K_s>2.3$ vs projected distance for QSOs (Open triangles) and optically obscured galaxies (Filled triangles). Error bars were estimated using jackknife re--sampling techniques.}
\label{ic}
\end{figure}

\section{Colour distribution of galaxies around QSOs and optically obscured sources}

In order to study the neighbour population of galaxies in these environments, we calculate the colour distribution of neighbouring galaxies according to the different target distances.
In Figure 3, we can see the percentage of DRGs respect to $15.5<K_s<19.5$ galaxies vs projected distance for two different target samples. QSO targets are represented with open triangles and optically obscured galaxies are represent with filled triangles. Error bars were estimated using $jackknife$ re--sampling techniques \citep{efron}. As can be seen, in the vicinity of these targets ($r < 0.3$ Mpc h$^{-1}$) the distribution of colours of galaxies is significantly different. Neighbouring galaxies close to QSOs tend to be bluer than galaxies in optically obscured source environments. For these objects, the fraction of galaxies with $J-K_s>2.3$ decreases from 20\% to 5\%.

\section{QSO-DRGs and OBSCURED-DRGs CROSS-CORRELATION ANALYSIS}
In this section we analyse the relative spatial clustering of DRGs and QSOs and optically obscured galaxies.
In order to obtain the cross--correlation length $r_0$, we first determine the projected cross--correlation function using QSOs as targets and the DRG population in their fields as tracer galaxies. We have done the same computation using optically obscured galaxies as targets and DRGs as tracer galaxies. 

We use the Peebles estimator \citep{peebles80} of the projected cross-correlation function:

\begin{equation}
\omega(\sigma)=\frac{n_{R}}{n_{G}}\frac{QG(\sigma)}{QR(\sigma)}-1,
\end{equation}
where $n_G$ and $n_R$ are the numbers of DRGs in the sample and in a random sample respectively, $QG(\sigma)$ is the number of real QSO-DRGs (or OBS-DRGs, using obscured galaxy targets) pairs separated by a projected distance in the range $\sigma$, $\sigma+\delta \sigma$,
and $QR(\sigma)$ are the corresponding pairs when considering the random
galaxy sample.

\begin{figure}
\includegraphics[width=80mm]{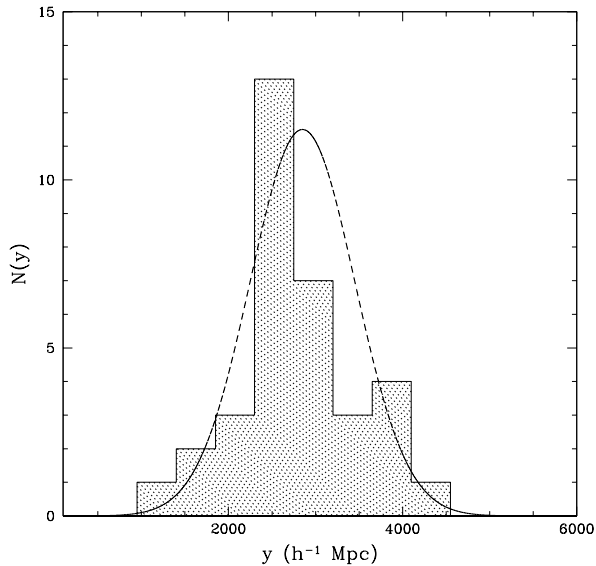}
\caption{Distance distribution of $K_s<19.5$ DRGs selected from the GOODS survey \citep{graziana} with photometric redshift measurements. Dashed lines shows the selection function obtained by fitting a Gaussian function.}
\label{ic}
\end{figure}

We have estimated the cross-correlation length taking into account the
selection function of the galaxy survey as a function of the distance
to the targets (QSOs or optically obscured galaxies, respectively).
Assuming a power-law model for the real space cross correlation
function it is found:

\begin{equation}
\omega(\sigma)=B \sqrt{\pi} \frac{\Gamma \left[(\gamma-1)/2\right] }{\Gamma(\gamma/2)}\frac{r_{0}^{\gamma}}{\sigma^{\gamma-1}},
\label{lil}
\end{equation}
where the constant $B$ depends on the differences of the selection
function of the different targets and tracers.

When the distance of these targets are known, the constant $B$ is given by \citep{lilje}:
 
\begin{equation}
B=\frac{\sum_{i} N(y_i)} {\sum_{i}\frac{1}{y_{i}^2}\int_{0}^{\infty}%
N(x)x^2dx},  
\label{b}
\end{equation}

where $N(y_i)$ is the selection function of the galaxy survey, $y_i$ is the
distance to target $i$ and the sum extends over all QSO (or obscured source) targets in the sample.

We model the selection function using the distance distribution of DRGs in the GOODS Survey (\citep{grazianb}) for galaxies with photometric redshifts and $K_s$ magnitudes $< 19.5$.
We convert measured redshifts to comoving distances along the line of sight assuming a flat $\Lambda$CDM model Universe with the same parameters quoted in the Introduction.
In Figure 4 we show the distance distribution of DRGs in the GOODS area with $K_s<19.5$ and with photometric redshift determinations. We model the distance distribution with a Gaussian function centred at $y$=2850 Mpc h$^{-1}$ and with a standard deviation of $\sigma_y$= 600 Mpc h$^{-1}$.

\begin{figure}
\includegraphics[width=80mm]{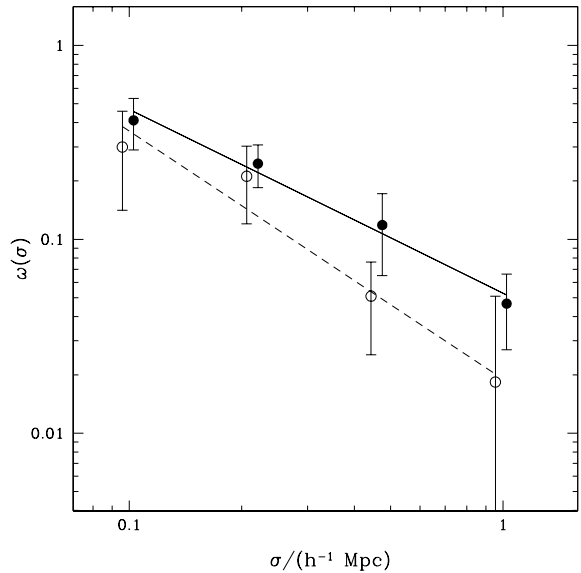}
\caption{Projected cross--correlation function between QSOs and DRGs in the redshift range $1\leq z\leq2$ (Filled circles) and between optically obscured galaxies with $z\sim2.2$ and DRGs with $K_s<19.5$ (Open circles). Error bars correspond to 1 $\sigma$ uncertainty estimated using the jackknife technique \citep{efron}.}
\label{ic}
\end{figure}

In Figure 5 we show the obtained projected cross--correlation function between QSOs and DRGs in the redshift range $1\leq z\leq2$ (Filled circles) and between optically obscured galaxies with $z\sim2.2$ and DRGs with $K_s<19.5$ (open circles). The error bars were estimated using the $jackknife$ technique \citep{efron}.
The corresponding cross--correlation length derived for the QSO sample targets is $r_0$=$5.4\pm1.6$ Mpc h$^{-1}$ and a slope of $\gamma$=$1.94\pm0.10$. Using the optically obscured galaxy sample as targets we find  $r_0$=$8.9\pm1.4$ Mpc h$^{-1}$ and a slope of $\gamma$=$2.27\pm0.20$. 
We tested the accuracy of this result by varying the $N(y_i)$ distribution over a reasonable 
range of values for $\bar y$ and $\sigma_y$. We find that the calculated $r_{0}$ is only weakly 
dependent on the $\bar y$ used, and is only affected at the 10\% level.

\section{Summary and discussions}

We have analysed the QSOs and optically obscured sources environment using the DRG photometric technique ($J-K_s>2.3$).
The sample of QSOs at $1\leq z\leq 2$ were selected from the Sloan Digital Sky Survey and the optically faint sources sample were detected in the mid-infrared with the Spitzer Space Telescope at z$\sim 2.2$ in the Flamingos Extragalactic Survey (FLAMEX).
We find that there are significant differences in the environment and galaxy properties around these two different targets. 
We find that neighbouring galaxies next to QSOs tend to be bluer than galaxies in optically obscured source environments.
The fraction of galaxies with $J-K_s>2.3$ in the vicinity ($r_p<0.3$ Mpc $h^{-1}$) of optically obscured sources decreases from 20\% to 5\%.
 These results are consistent with those of \citet{coil} who find that the local environment of $0.7<z<1.4$ QSOs (using the $3^{rd}$--nearest--neighbour surface density of surrounding DEEP2 galaxies)  has a similar overdensity than that of blue galaxies, differing from red galaxy population at 2$\sigma$ significance.
We also present results on the cross--correlation function of DRGs around QSOs and optically obscured sources. 
The corresponding cross--correlation length derived for the QSO and the DRGs is $r_0$=$5.4\pm1.6$ Mpc h$^{-1}$ with a slope of $\gamma$=$1.94\pm0.10$. 
Using the optically obscured galaxy sample as targets we find  $r_0$=$8.9\pm1.4$ Mpc h$^{-1}$ and a slope of $\gamma$=$2.27\pm0.20$. 
We point out that the results of Sections 4 and 5 consistently indicate that obscured sources are located in higher density environment compared to QSOs, as shown by both, the colour distribution and the cross-correlation analysis.  

It is possible that the optically obscured sources population are not rare objects.
Using a QSO sample in the full Bo{\"o}tes field, \citet{weedman} found that unobscured, classical type 1 QSOs and optically obscured sources with similar redshifts are similar in number.  
As noted by \citet{weedman}, sources selected at 24 $\mu$m, which are optically faint, are dominated by dusty sources with spectroscopic indicators of an obscured AGN rather than a star-burst. \citet{houck} found that these sources represent ULIRGS powered by a AGN and they have bolometric luminosities exceeding $10^{13}$ M$\sun$, if bolometric luminosities scale with mid--infrared luminosities as in the templates which fits the spectra. This value is similar to those found for bright QSOs at $z\sim2$ \citep{hop}.
\citet{elli} have showed that luminous QSOs in rich galaxy cluster environments evolves much more quickly than those in poor environments which are present at all redshifts. It is possible that the QSO activity in optical obscured sources have vanished as a consequence of the rich environment where they are located or the luminous central engine is heavily obscured by the dust. 

Few studies exists in the Literature about the spatial clustering of DRGs.
\citet{foucaud} found a large correlation length of $r_0\sim12$ $h^{-1}$ Mpc for a sample of bright DRGs ($K_{AB}\sim$20.7, or $K_{Vega}$$\sim$ 18.8) in a small area of the Early data release of the UKIDSS survey.
\citet{graziana} found $r_0=7.41_{-4.84}^{+3.45}$ $h^{-1}$ Mpc for DRGs with $1 < z < 2$ and $K_{AB}<$22 ($K_{Vega}< 20$). The larger correlation length found by \citet{foucaud} could be interpreted as evidence for luminosity segregation, where the most luminous DRGs are more strongly clustered. 
If we compare our values obtained in the cross-correlation function analysis with the mean value obtained for the DRGs in the field by \citet{graziana}, our results suggest that the sample of unobscured QSOs at high redshift analysed tend to be located in typical environment of DRGs with a transient phase in the evolution of normal galaxies. However, our cross-correlation results indicate that optically obscured sources probably represent massive galaxies located in rich environment of DRGs at high redshifts.

\section{Acknowledgements}
We are thankful to the referee Dr. Michael Strauss for his careful reading of the manuscript and a number of comments, which improved the paper.

This paper used public catalogues form the Flamingos Extragalactic Survey,
with support from NSF grant AST97-31180 and Kitt Peak National Observatory 
This work was partially supported by the
Consejo Nacional de Investigaciones Cient\'{\i}ficas y T\'ecnicas (CONICET)
 and the Secretar\'ia de Ciencia y
T\'ecnica de la Universidad Nacional de C\'ordoba.

This research has made use of the NASA/IPAC Extragalactic Database (NED), which is operated by the Jet Propulsion Laboratory, California Institute of Technology, under contract with the National Aeronautics and Space Administration.

Funding for the SDSS and SDSS-II has been provided by the Alfred P. Sloan Foundation, the Participating Institutions, the National Science Foundation, the U.S. Department of Energy, the National Aeronautics and Space Administration, the Japanese Monbukagakusho, the Max Planck Society, and the Higher Education Funding Council for England.
The SDSS Web Site is http://www.sdss.org/. The SDSS is managed by the Astrophysical Research Consortium for the Participating Institutions. The Participating Institutions are the American Museum of Natural History, Astrophysical Institute Potsdam, University of Basel, Cambridge University, Case Western Reserve University, University of Chicago, Drexel University, Fermilab, the Institute for Advanced Study, the Japan Participation Group, Johns Hopkins University, the Joint Institute for Nuclear Astrophysics, the Kavli Institute for Particle Astrophysics and Cosmology, the Korean Scientist Group, the Chinese Academy of Sciences (LAMOST), Los Alamos National Laboratory, the Max-Planck-Institute for Astronomy (MPIA), The Max-Planck-Institute for Astrophysics (MPA), New Mexico State University, Ohio State University, University of Pittsburgh, University of Portsmouth, Princeton University, the United States Naval Observatory, and the University of Washington.

{}

\setcounter{table}{1}
\begin{onecolumn}
\label{t0}
\begin{tabular}{lc}
\hline
(1)&(2)\\
Name  &redshift  \\
\hline
SST24 1435203.99+330706.8 &  2.59$\pm$0.34  \\
SST24 J142804.12+332135.2 &  2.34$\pm$0.28  \\
SST24 J143358.00+332607.1 &  1.96$\pm$0.34   \\
SST24 J143001.91+334538.4 &  2.46$\pm$0.20   \\
SST24 J143251.82+333536.3 &  1.78$\pm$0.14   \\
SST24 J143520.75+340418.2 &  2.08$\pm$0.21  \\
SST24 J143026.05+331516.4 &  1.90$^*$          \\
SST24 J143429.56+343633.1 &  2.00$^*$       \\
\hline
\end{tabular}

Table 2. Optically obscured sources sample characteristics. SST24 source name derives from discovery with the MIPS 24 $\mu$m images, 
redshifts derived from strong ($^*$ weak) silicate absorption features.
\end{onecolumn}

\setcounter{table}{2}
\begin{onecolumn}
\label{t0}
\begin{tabular}{lc}
\hline
(1)&(2)\\
Name  &redshift \\
\hline
 SDSS J143106.78+340910.9     &  1.098000   \\
 SDSS J143132.13+341417.3     &  1.039840    \\
 SDSS J143331.80+341532.8     &  0.957354   \\
 SDSS J143421.33+340446.9     &  1.956500   \\
 SDSS J142912.88+340959.1     &  2.229540    \\
 SDSS J143307.89+342315.9     &  1.950550    \\
 SDSS J143201.75+343526.2     &  1.070690   \\
 RIXOS F110--50               &  1.335000  \\
 SDSS J142744.44+333828.7     &  1.237000   \\
 SDSS J143605.08+334242.6     &  1.983830    \\
 SDSS J143628.09+335524.3     &  0.903084     \\
 SDSS J143543.72+342906.4     &  2.547330   \\
 SDSS J143627.79+343416.8     &  1.883490   \\
\hline
\end{tabular}\\
Table 3. QSOs sample characteristics. Designation in IAU format and spectroscopic redshift.
\end{onecolumn}

\end{document}